\documentclass[twocolumn,aps,prl,showpacs,floatfix,,superscriptaddress]{revtex4}
\usepackage[latin9]{inputenc}
\usepackage{graphicx}
\usepackage{esint}

\makeatletter
\@ifundefined{textcolor}{}
{%
 \definecolor{BLACK}{gray}{0}
 \definecolor{WHITE}{gray}{1}
 \definecolor{RED}{rgb}{1,0,0}
 \definecolor{GREEN}{rgb}{0,1,0}
 \definecolor{BLUE}{rgb}{0,0,1}
 \definecolor{CYAN}{cmyk}{1,0,0,0}
 \definecolor{MAGENTA}{cmyk}{0,1,0,0}
 \definecolor{YELLOW}{cmyk}{0,0,1,0}
}

\usepackage{bm}\usepackage{float}\usepackage{afterpage}

\makeatother

\begin{document}

\title{Bosonic edge states in gapped honeycomb lattices}

\author{Huaiming Guo}
\thanks{hmguo@buaa.edu.cn}
\affiliation{Department of Physics, Beihang University, Beijing, 100191, China}

\author{Yuekun Niu}
\affiliation{Department of Physics, Beijing Normal University, Beijing, 100875, China}

\author{Shu Chen}
\affiliation{Beijing National
Laboratory for Condensed Matter Physics, Institute of Physics,
Chinese Academy of Sciences, Beijing 100190, China}
\affiliation{Collaborative Innovation Center of Quantum Matter, Beijing, China}

\author{Shiping Feng}
\affiliation{Department of Physics, Beijing Normal University, Beijing, 100875, China}

\begin{abstract}
Using quantum Monte Carlo simulations of bosons in gapped honeycomb lattices, we prove the existence of bosonic edge states. For single-layer honeycomb lattices, bosonic edge states can be created, cross the gap, and merge into bulk states using an on-site potential applied to the outermost sites of the boundary. On the bilayer honeycomb lattice, bosonic edge states traversing the gap at half filling are demonstrated. The topological origin of the bosonic edge states is discussed in terms of the pseudo-Berry curvature. The results will simulate experimental studies of these exotic bosonic edge states using ultracold bosons trapped in honeycomb optical lattices.
\end{abstract}


\maketitle

\textit{Introduction.-} Many important phenomena in condensed matter physics share the same intriguing feature, namely, the presence of edge states in the gap. Well-known examples include the quantum Hall effect (QHE), topological insulators, and graphene systems \cite{ti1,ti2,ti3,graphene}. The appearance of an edge state is the consequence of the non-trivial topological properties of the bulk band. Edge states have been created in various systems such as quantum wells and graphene \cite{tiedge1,qwelledge2,grapheneedge3,photonicedge4}; the existence of edge states constitutes a central motivation for studying these kinds of materials. Edge states are perfectly conducting channels and play an important role in electronic transport \cite{hgte}. Edge states can be engineered to create one-dimensional topological superconductors, which support localized Majorana fermions \cite{majorana}.

Generalization of the above-described fermion phenomena in systems of bosons has become an active area of research. Various kinds of bosonic Dirac materials have been proposed \cite{sug1,sug2,sug3,sug4,sug5}. A general possibility of the existence of the bosonic QHE based on effective field theory has been proposed \cite{qft}.  Using direct analogs of the fractional QHE (FQHE) in a topological flat band \cite{flat,ffqhe}, evidence of bosonic FQHE has been demonstrated in a simple lattice model \cite{bfqhe}. However, this simple replacement in a Chern insulator does not generate the expected bosonic QHE and the bosonic phase dominated by spin-orbit coupling in the Haldane model becomes a chiral superfluid \cite{bosonichaldane1,bosonichaldane2,bosonichaldane3}. The realization of bosonic QHE in a simple lattice is achieved by combining correlated hopping and a background gauge field or the use of the lowest band of an optical flux lattice \cite{biqhe1,biqhe2,biqhe3}. A hallmark of bosonic QHE is that it supports gapless edge states. Until now, bosonic edge modes have only been probed using entanglement spectra; direct demonstration of these modes has yet to be performed.

It is well known that graphene systems exhibit edge states under certain boundary conditions, although they have a different physical origin from those in QHE and topological insulators.
Free electrons in a honeycomb lattice mimic Dirac fermions characterized by the Berry phase $\pi$, which results in a one-dimensional flat band connecting the two Dirac points on zigzag or bearded edges \cite{berry,1dflatband}. Although the Dirac fermions can be experimentally realized by loading ultracold fermions into honeycomb optical lattices \cite{ZhuSL}, it is not clear whether a bosonic system can exhibit similar physical phenomena, as bosons obey a different statistical rule and the band structure collapses, even for hardcore bosons. Therefore, it is useful to determine whether the topological property remains for bosons in honeycomb lattice, and if so, whether bosonic edge states can be achieved.

In this paper, we study extended Bose-Hubbard models of single-layer and bilayer honeycomb lattices. For the single-layer case, a $\rho=\frac{1}{2}$ charge-density-wave (CDW) insulator is obtained using a staggered sublattice potential. After applying an on-site potential at the boundary, the existence of bosonic edge states is explicitly demonstrated through the appearance of states with finite superfluid density in the gap. The bosonic edge states can be induced, via the applied potential, to cross the gap and merge into the bulk states below the gap. The edge-like nature of the system is confirmed by the distribution of the edge state, which is primarily found near the boundary. We also find that the results remain valid for various CDW insulators formed by interacting softcore bosons. Finally, we study a bilayer honeycomb lattice in which a gap is opened by applying opposite potentials to the two layers. A bosonic edge state traversing the gap is demonstrated. The results will of interest in cold-atom experiments; observation of the bosonic edge state is possible based on currently available experimental setups.

\begin{figure}[htbp]
\centering
\includegraphics[width=7cm]{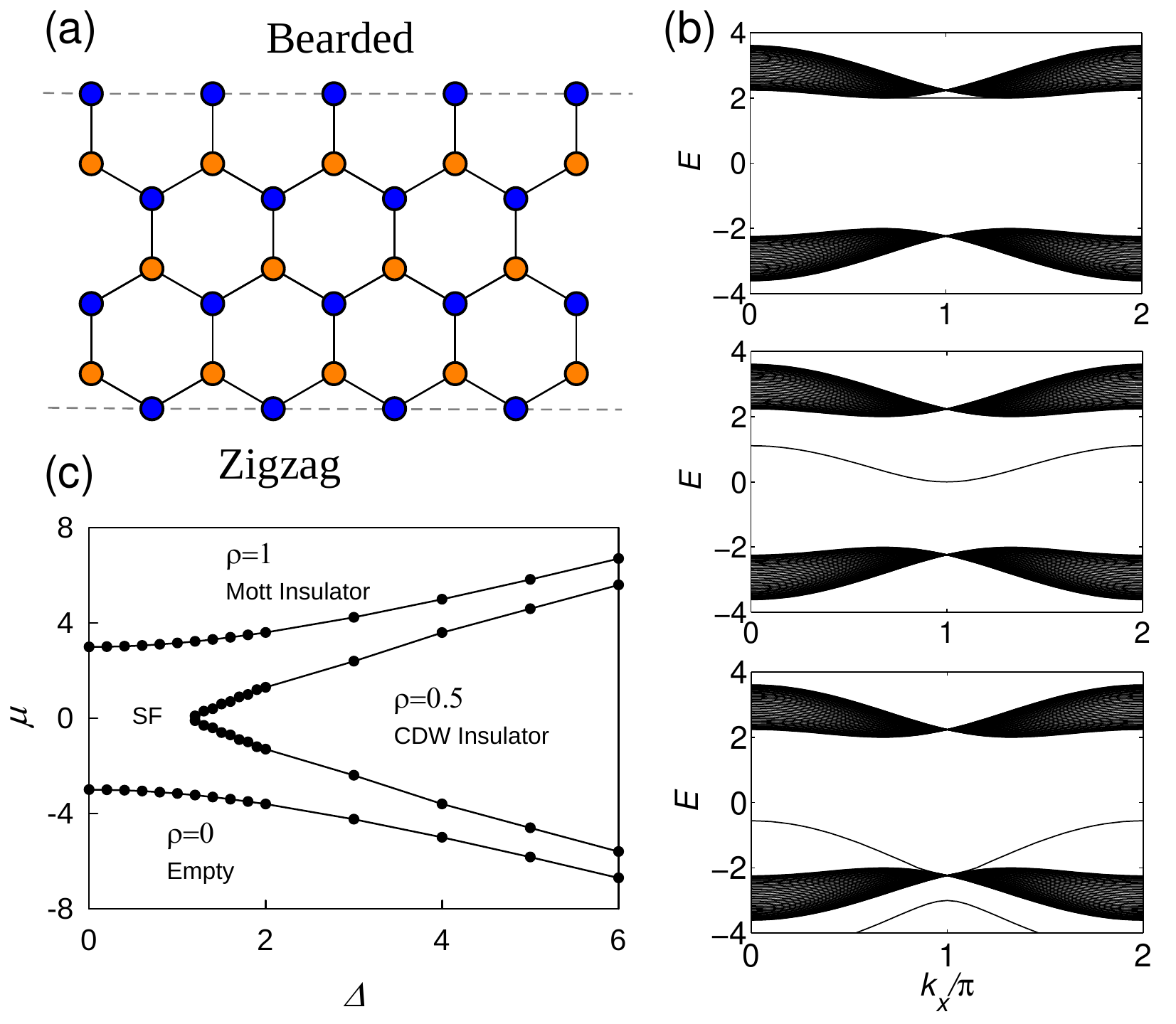}
\caption{(Color online) (a) Schematic illustration of the zigzag and bearded edges on honeycomb lattices. (b) Band structures with different on-site potentials of the outermost sites on zigzag edges: $U' = 0$ (upper), $U' = -2$ (middle) and $U' = -5$ (lower). Here, the staggered sublattice potential is $\Delta = 2$.  (c) Phase diagram of the Hamiltonian in Eq. (\ref{eq1}), which contains three insulating regions, corresponding to zero, half, and full fillings, and a superfluid phase.}\label{fig1}
\end{figure}

\textit{Model and method.-} First, we consider the extended Bose-Hubbard model of a single-layer honeycomb lattice, which is described by the following Hamiltonian:
\begin{eqnarray}\label{eq1}
\hat{H}=-t\sum_{\langle i,j\rangle} (\hat{b}^{\dagger}_{i}\hat{b}_{j}+\hat{b}^{\dagger}_{j}\hat{b}_{i}) +\sum_{i}U_{i}\hat{n}_{i}-\mu\sum_{i} \hat n_{i}.
\end{eqnarray}
Here, $\langle i,j\rangle$ runs over the nearest-neighbor (NN) pairs. $\hat{b}^{\dagger}_{i} (\hat{b}_{i})$ denotes the hardcore boson creation (annihilation) operator, which obeys the commutation relation $[\hat{b}_i,\hat{b}^{\dagger}_j]=0$ for $i\neq j$, abd the anti-commutation relation $\{\hat{b}_i, \hat{b}^{\dagger}_i\}=1$ at the same site. $\hat{n}_{i}=\hat{b}^{\dagger}_{i}\hat{b}_{i}$ is the local density operator. $\mu$ is the chemical potential, which controls the filling of the lattice. $U_{i}$ is a staggered sublattice potential: $U_{i}=\Delta$ for sublattice A and $U_{i}=-\Delta$ for sublattice B.

For fermions with the same form of Hamiltonian, the energy spectrum contains two branches,
\begin{eqnarray}\label{eq2}
E=\pm \sqrt{\Delta^2+t^2(2\cos \frac{\sqrt{3}}{2}k_{x}+\cos \frac{3}{2}k_{y})^2+t^2\sin^2 \frac{3}{2}k_{y}},
\end{eqnarray}
where $(k_{x},k_{y})$ are momenta. The spectrum is symmetric about $E = 0$ and has a gap of $2\Delta$ at two inequivalent Dirac points $K_{\pm}=(\pm \frac{4\pi}{3\sqrt{3}},0)$ located at the corners of the hexagonal Brillouin zone. The energy of the bottom of the spectrum is $E_{B} = -\sqrt{\Delta^2+9t^2}$, which is the chemical potential when bosons begin to fill the the lattice in the grand canonical ensemble. It is well known that flat edge states connecting the two Dirac points appear in the region $k_x\in [\frac{2}{3}\pi, \frac{4}{3}\pi]$ for a zigzag edge and in the complementary region for a bearded edge [see Fig.\ref{fig1}(b)]. Additionally, because we have a strictly localized state at the edge for $k_x = \pi$, the edge states can be controlled via the potentials applied to the boundary \cite{wyao,semenoff}.

To study the Hamiltonian in Eq. (\ref{eq1}), we employ the stochastic series expansion quantum Monte Carlo (QMC) method with directed loop updates, which is realized using the ALPS library \cite{sandvik,alps}. To characterize the different phases, we calculate the density difference of the two sublattices and the superfluid density \cite{wessel,wen},
\begin{eqnarray}\label{eq3}
\rho_{A-B}=|\rho_A-\rho_B|, \rho_s=\frac{\langle W^2\rangle}{2\beta t},
\end{eqnarray}
where $\rho_A$ and $\rho_B$ are the boson density for the $A$ and $B$ sublattices, respectively; $W$ is the winding number of the world line; and $\beta$ is the inverse temperature. A superfluid phase is characterized by $\rho_{A-B}=0$ and $\rho_s\neq 0$, whereas a solid phase is characterized by $\rho_s=0$. In the following section, we focus on zigzag edges; similar results are found for bearded edges.

\textit{Bosonic edge states in CDW insulators.-} First, the phase diagram of the Hamiltonian in Eq. (\ref{eq1}) is mapped.
In this map, it is useful to consider the atomic limit $t=0$. Whether a boson can be added to the $i$-th site is determined by the energy difference $\Delta E=-\mu+U_{i}$. If $\Delta E<0$, the total energy is lowered and each site is occupied by one hardcore boson. Thus, two lines $\mu=\pm \Delta$ separate the different insulators. There are two kinds of insulators in the phase diagram: a Mott insulator with a uniform density $\rho=1$ and a CDW insulator with a density profile reflecting the staggered superlattice potential and an average density $\rho=\frac{1}{2}$. After NN hopping is turned on, the phase diagram, shown in Fig.\ref{fig1} (c), is obtained from the QMC simulation. For $\mu<E_{B}$, the system is empty, whereas for $\mu>-E_{B}$, it is fully occupied. The CDW insulator in the atomic limit persists; however, its boundary is deformed and an incommensurate superfluid region appears between the commensurate insulators. The phase transition between the superfluid and insulator is continuous.

\begin{figure}[htbp]
\centering
\includegraphics[width=8cm]{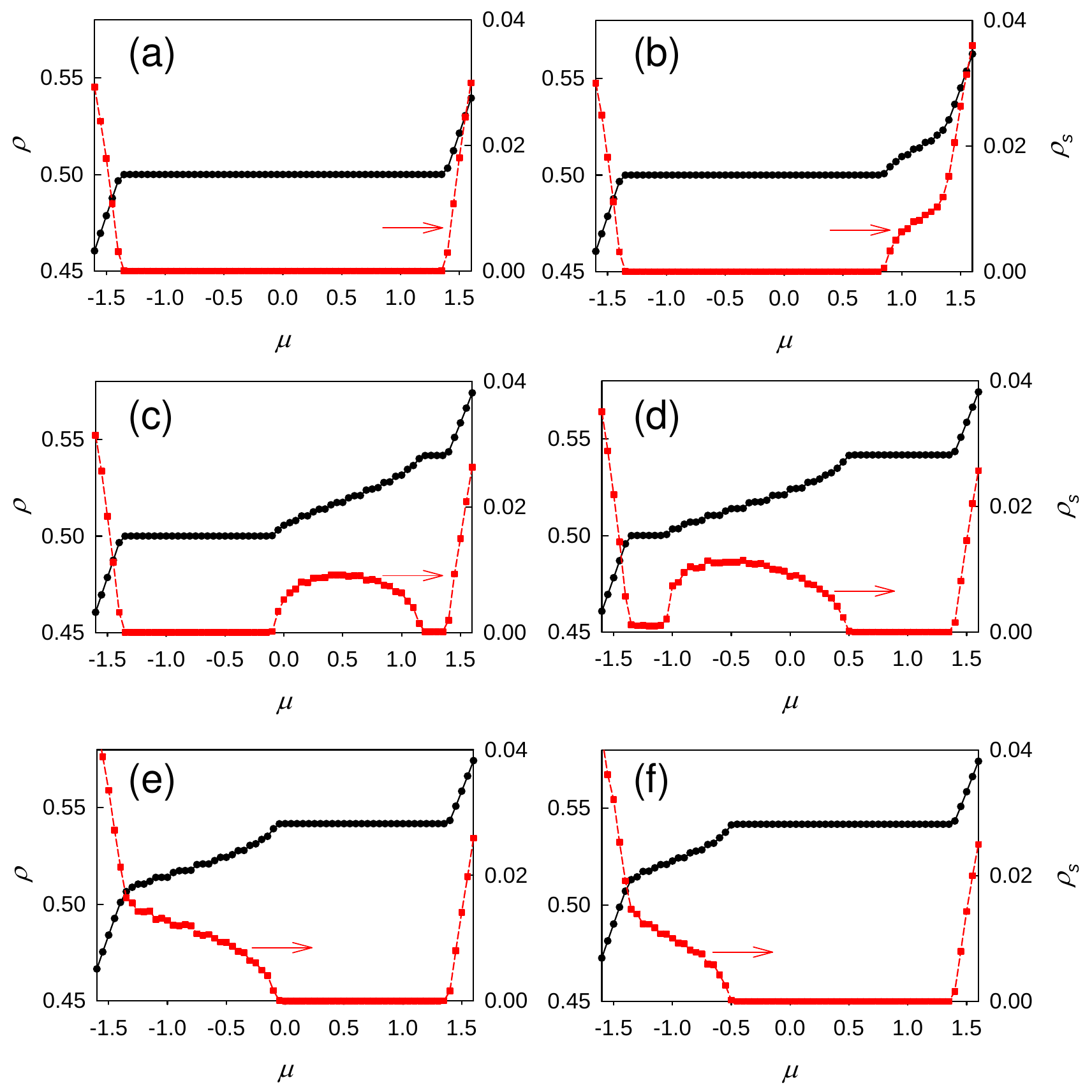}
\caption{(Color online) The average density and superfluid of the Hamiltonian in Eq. (1) with zigzag edges. The on-site energies of the outermost sites on the zigzag edges are tuned to different values: (a) $U'=0$, (b) $U'=-1$, (c) $U'=-2$, (d) $U'=-3$, (e) $U'=-4$, and (f) $U'=-5$. The staggered sublattice potential is $\Delta=2$ and the lattice size is $L=12$. The black (red) curves represent $\rho (\rho_s)$.}\label{fig2}
\end{figure}

Next, we study the edge states on zigzag edges in the $\rho=\frac{1}{2}$ CDW insulator. The geometry of the system considered is a strip with zigzag edges and a periodic boundary condition in the $x$ direction. In the presence of the edge, the system remains gapped; the gap size is not significantly affected, as reflected by the $\rho,\rho_s \sim \mu$ curves in Fig.\ref{fig2}(a). However, the superfluid density is reduced to roughly half, which can be understood as being caused by the winding number across the open boundary vanishing. The edge state appears after an on-site potential on the outermost sites is applied. As shown in Fig.\ref{fig2}(b), once the potential is applied, an edge state extending into the gap from its upper border appears. This situation is similar to the fermionic case, in which the dispersion bends downward into the gap when the same potential is applied. Additionally, the corresponding superfluid density has a finite value, implying that the edge state forms a superfluid along the edge. The displacement of the bottom of the edge state is roughly equal to the value of the applied potential, implying that the corresponding state is completely localized on the outermost sites. As the potential further increases [Fig.\ref{fig2}(c)-(f)], the edge state crosses the gap and tends to merge into the region below the gap. Note that for a finite region of the potential, the edge state is completely within the gap. This situation is caused by the large gap at $\Delta=2$; the same behavior occurs for the fermionic system [see Fig.\ref{fig1}(b)]. However, for smaller values of $\Delta$, the edge state can traverse the gap.

\begin{figure}[htbp]
\centering
\includegraphics[width=8cm]{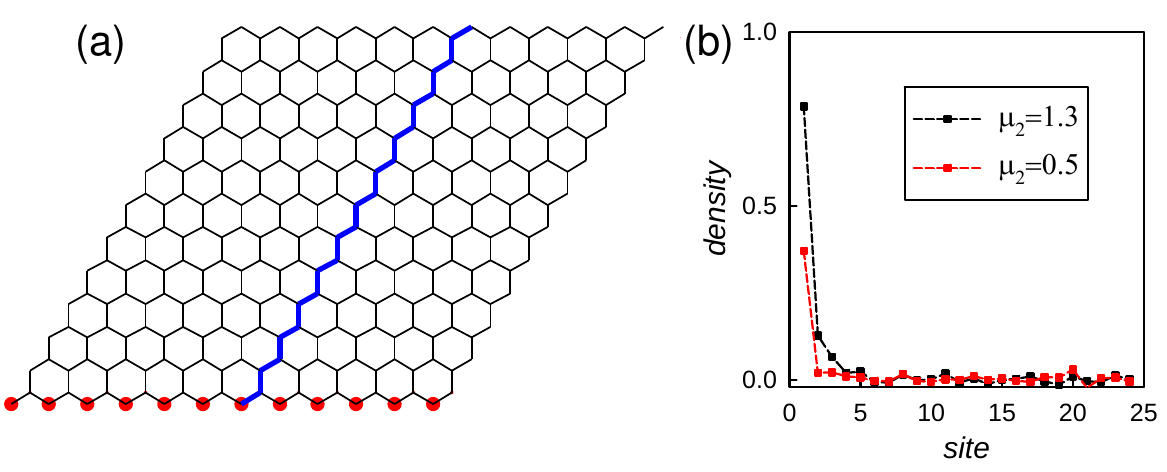}
\caption{(a) (Color online) The distribution of the edge state in Fig.\ref{fig2}(c). The value of the number of particles is denoted by the size of the red circle. (b) The distribution of the edge state in sites along the blue line in (a). In (b), the red curve is the distribution along the line when the edge state is partially filled. Here $\mu_1=-0.2$ corresponds to the lower plateau. }\label{fig3}
\end{figure}

When the edge state is completely in the gap [see Fig.\ref{fig2}(c) and (d)], there are two plateaus with vanishing superfluid in the $\rho \sim\mu$ curve; these plateaus reflect the gaps between the edge state and the bulk states. The value of the lower plateau is $\rho_1=0.5$, whereas the value of the upper plateau is $\rho_2=0.5417$ (corresponding to the lattice size $L=12$ used in the simulations). The number of hardcore bosons in the edge state is directly calculated using $\delta n=2L^2(\rho_2-\rho_1)\approx L$, which is equal to the number of outermost sites of the zigzag edge and remains valid for larger sizes. To further confirm the edge nature of the in-gap state, we calculate the distribution of hardcore bosons along the blue line in Fig. \ref{fig3}(a), which is defined as: $\delta n_{i}=n_{i}(\mu_{2})-n_{i}(\mu_{1})$ where $n_{i}(\mu_{\alpha}) (\alpha=1,2)$ is the local density at the chemical potential $\mu_{\alpha}$ corresponding to the average density $\rho_i$. As shown in Fig. 3(b),  the distribution of the edge state in Fig. \ref{fig2}(c), which is $\delta n_i$ between the plateaus, is distributed primarily near the boundary.

\begin{figure}[htbp]
\centering
\includegraphics[width=6cm]{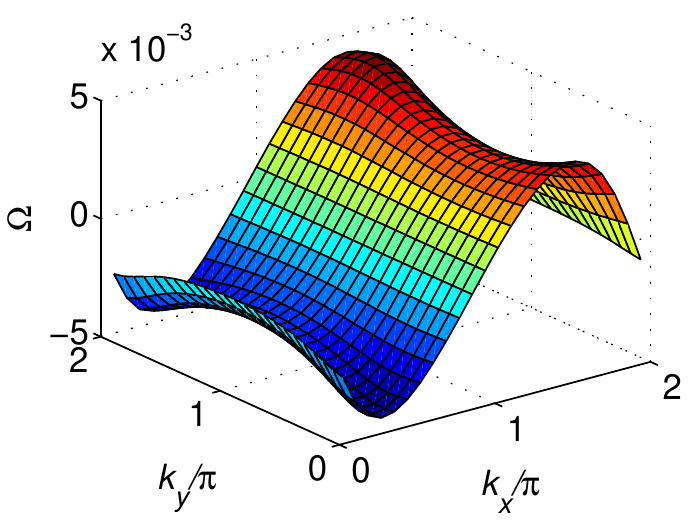}
\caption{(Color online) Pseudo-Berry curvature of the Hamiltonian in Eq. (\ref{eq1}) obtained from QMC simulation. Here, $\Delta=2$ and the system is a $\rho=\frac{1}{2}$ CDW insulator. }\label{fig4}
\end{figure}

Usually, the appearance of an edge state is a consequence of the non-trivial topological property of bulk system. Although band structures collapse for bosonic systems, a pseudo-Berry curvature could be defined using the equal-time Green function, which can be obtained from the QMC calculation \cite{bberry, bdm}. In Fig. \ref{fig4}, we show the pseudo-Berry curvature for the above $\rho=\frac{1}{2}$ CDW insulator. Although no peaks form at the same Dirac points as in the fermionic counterpart, the pseudo-Berry curvature develops two non-zero regions with different signs. The bosonic edge states may appear to connect the two regions. Thus, when the on-site potential is applied to the outermost sites, they can be tuned into the gap.

\begin{figure}[htbp]
\centering
\includegraphics[width=8cm]{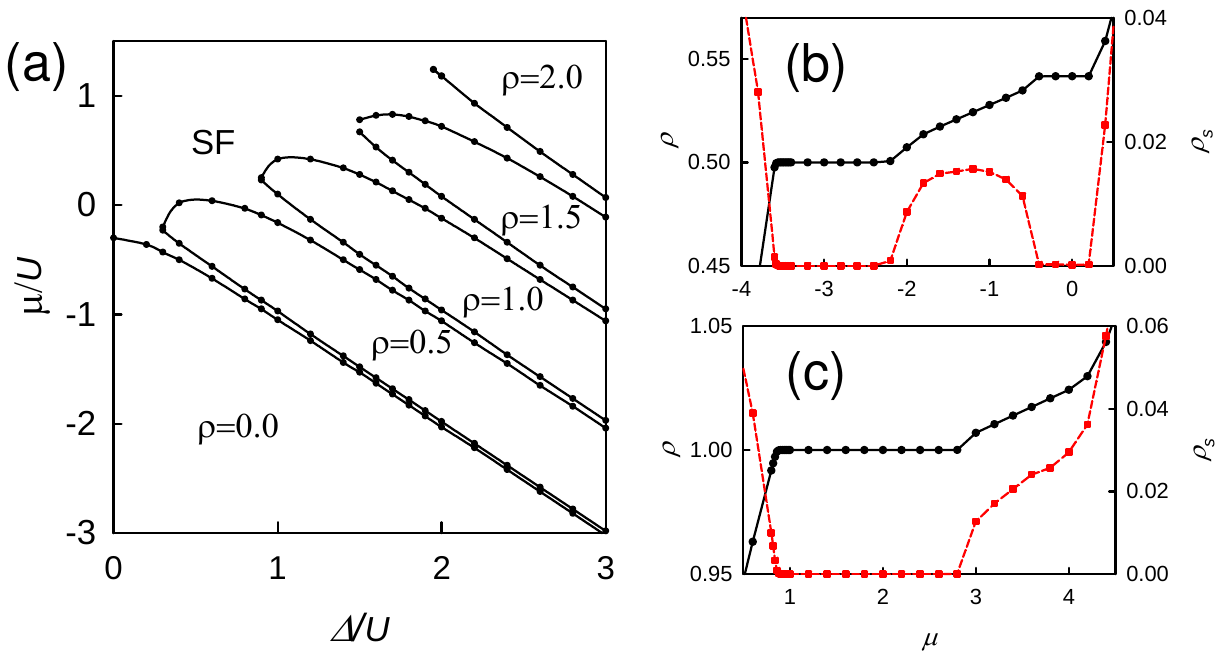}
\caption{(Color online) (a) The phase diagram of softcore bosons for $U=10$ in the $(\Delta, \mu)$ plane. The average density and superfluid of the bosonic edge states: (b) $\rho=\frac{1}{2}$ and (c) $\rho=1$. Here, the applied potential on the boundary is $ U'=-5$. The staggered sublattice potential is $\Delta=4$ in (b) and $\Delta=10$ in (c).}\label{fig5}
\end{figure}

Perviously, we have established the existence of edge state of hardcore bosons in a honeycomb lattice with a zigzag edge and have shown that they can be tuned by applying an on-site potential to the outermost sites. In fact, the results remain valid for interacting softcore bosons. When subjected to a staggered sublattice potential, the softcore Bose-Hubbard model produces various CDW insulators at half-integer and full-integer fillings [see Fig. \ref{fig5}(a)]. These CDW insulators are adiabatically connected to those in the atomic limit with one sublattice occupied. Under a boundary potential, similar evolutions of edge states are found. We show the bosonic edge states at $U'=-5$ in two typical CDW insulators with $\rho=\frac{1}{2}$ and $\rho=1$. Compared to the case of hardcore bosons, larger potentials are needed to tune the edge states.

\begin{figure}[htbp]
\centering
\includegraphics[width=8cm]{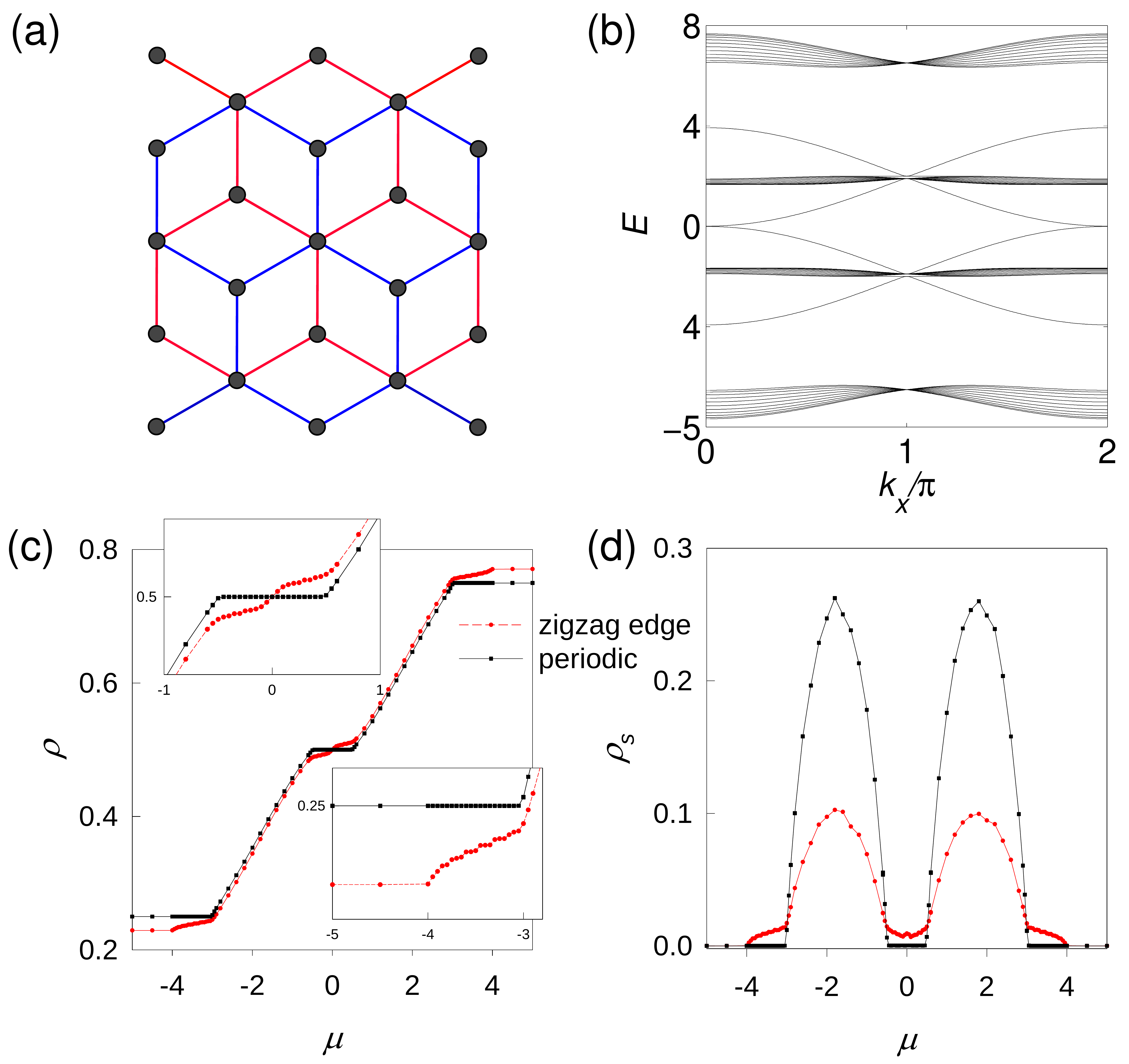}
\caption{(Color online) (a) Lattice structure of a bilayer honeycomb lattice with Bernal stacking. (b) Band structure of the gapped bilayer honeycomb lattice with zigzag edge at $t_{\bot}=-6$ and $\Delta=2$. The average density (c) and superfluid (d) as a function of $\mu$ in the presence and absence of edges. The inset in (c) shows enlarged images of the plateau at half filling and the lower plateau at quarter filling.}\label{fig6}
\end{figure}

\textit{Bilayer honeycomb lattice.-} Finally, we study the edge state of hardcore bosons on the bilayer honeycomb lattice with Bernal stacking. It is useful to first analyze the band structure from the bilayer Hamiltonian \cite{bilayer},
\begin{eqnarray}\label{eq3}
H=\left(
    \begin{array}{cccc}
      \Delta & h_{+} & 0 & t_{\bot} \\
      h_{-} & \Delta & 0 & 0 \\
      0 & 0 & -\Delta & h_{+} \\
      t_{\bot} & 0 & h_{-} & -\Delta \\
    \end{array}
  \right),
\end{eqnarray}
where $h_{\pm}=h_x\pm ih_y$ where $h_x=t[2\cos \frac{\sqrt 3}{2}k_x \cos \frac{1}{2}k_y+\cos k_y]$ and $h_y=t[2\cos \frac{\sqrt 3}{2}k_x \sin \frac{1}{2}k_y-\sin k_y]$, $t_{\bot}$ is the interlayer coupling, and $\Delta$ is an on-site potential with opposite sign on the two layers, which is included so as to open a gap. For large values of $\Delta$, the two layers are effectively decoupled and a gap $\sim 2\Delta-B_{w}$ ($B_{w}$ the band width) opens at half filling. For large $t_{\bot}$, a gap $\sim 2\Delta$ opens at half filling. When both $\Delta$ and $t_{\bot}$ are sufficiently large, gaps also open at $\frac{1}{4}$ and $\frac{3}{4}$ fillings. For insulators dominated by $t_{\bot}$, the edge state traversing the gap appears on the zigzag edge [see Fig. \ref{fig6}(b)] because of the topological charge of the bulk.

Next, we fill the above bilayer lattice with hardcore bosons. As shown in Fig. \ref{fig6}(c) and (d), bosonic insulating phases appear at $\rho=\frac{1}{4}, \frac{1}{2}, \frac{3}{4}$ fillings. The gaps corresponds to the widths of the plateaus in the $\rho \sim \mu$ curve and are reduced compared to those in the fermionic system. In the presence of the zigzag edge, the density plateau at half filling changes to a small slope, indicating the appearance of an in-gap edge state. Because there is no gap between the edge state and the bulk state, the bosons traverse the gap, which is similar behavior as in the fermionic case. In the region in which the edge state exists, the superfluid density is finite. Thus, the bosonic system is insulating in the bulk, but a superfluid flows along the edge.
Edge states also appear near the upper (lower) border of the gap at $\frac{1}{4}$ ($\frac{3}{4}$) filling. These states are the counterparts of the fermionic state in Fig. \ref{fig6}(b).

\textit{Conclusions and discussions.-} In summary, we studied the bosonic edge states in gapped honeycomb lattices; their existence is clearly demonstrated by the appearances of states with finite superfluid density in the gap. Whereas the bosonic edge state could be controlled by applying an on-site potential to the outermost sites of the boundary in a single-layer lattice, the bosons traverse the gap in a bilayer system. The results remain valid for interacting softcore bosons. The appearance of the bosonic edge states may be attributed to a non-zero pseudo-Berry curvature.

The results are closely related to ultracold atom experiments. Honeycomb optical lattices were realized using three interfering traveling laser beams \cite{honeycomb1,honeycomb2,honeycomb3,honeycomb4,honeycomb5}. The staggered sublattice potential can be precisely adjusted. A Bose-Einstein condensate has been investigated in a honeycomb lattice. Therefore, the underlying Hamiltonian studied here is directly engineered using existing experimental techniques. There are many proposals to engineer sharp boundaries and schemes to detect edge states in optical lattices \cite{boundary1,boundary2,boundary3,boundary4}. Additionally, much progress has been made in high resolution and single-atom detection of atoms in a lattice \cite{singleat1,singleat2,singleat3}. Therefore, the observation of the studied bosonic edge state should be possible based on currently available experimental setups.

\textit{Acknowledgments-}
The authors wish to thank G. Chen, M. Dolfi, B. Gremaud, S. Isokov, Y. Li, T. Mendes, R. Scalettar, M. Troyer, L. Wang, Y.C. Wang, and T. Ying for helpful discussions. H.G. is supported by NSFC under Grants No. 11274032 and No. 11104189. S.C. is supported by NSFC under Grants No. 11425419 and No. 11325417. S.F. is supported by funds from the Ministry of Science and Technology of China under Grant No. 2012CB821403, and NSFC under Grants No. 11274044 and No. 11574032.

\end{document}